\documentclass[twocolumn,10pt]{article}

\usepackage{graphicx}
\usepackage{amsmath}
\usepackage[english]{babel}
\usepackage{macros}

\begin{document}

\title{Security Policy Consistency}

\author{Carlos Ribeiro \hspace{1cm} Andr\'{e} Z\'{u}quete \hspace{1cm} Paulo Ferreira \hspace{1cm} Paulo Guedes\\
        {\small\{Carlos.Ribeiro,Andre.Zuquete,Paulo.Ferreira,Paulo.Guedes\}@inesc.pt}
        \\{\small IST / INESC}\\
        {\small Rua Alves Redol N$^{o}$9 1000 Lisboa} \\
        {\small Portugal} \\}
\date{}
\maketitle

\begin{abstract}

With the advent of wide security platforms able to express
simultaneously all the policies comprising an organization's
global security policy, the problem of inconsistencies within
security policies become harder and more relevant.

We have defined a tool based on the CHR language which is able to
detect several types of inconsistencies within and between
security policies and other specifications, namely workflow
specifications.

Although the problem of security conflicts has been addressed by
several authors, to our knowledge none has addressed the general
problem of security inconsistencies, on its several definitions
and target specifications.

\end{abstract}

\section{Introduction}
\label{sec:introduction}

Over the years several access control policies have been proposed
in the literature. Although these policies cover many different
situations and types of information, they are often considered in
isolation. Thus, they are not suitable for organizations with
complex structures, that manage simultaneously several types of
information, thus requiring the simultaneous use of different
access control policies. Moreover, policies are often scattered
over different environments, making understanding and managing of
global policies much more difficult.

Recently there has been a considerable interest in environments
that support multiple and complex access control policies,
\cite{Bertino:1999:CSFW,Jajodia:1997:UFE,SSP92*33,ribeiro99a}.
The goal of these environments is to provide support for the
definition of all the policies that makes up the global security
policy of an organization into one single platform, thus
simplifying management and consistency maintenance.

Some of these environments provide mechanisms to solve potential
conflicts between contradictory policies. Some of these
mechanisms use special ad-hoc rules to decide upon the
acceptability of an action whenever a conflict arises
\cite{Jajodia:1997:UFE}; others use properties such as
``authorship", ``specificity" and ``recency" of security policies
to decide on their priority
\cite{Bertino:1999:CSFW,Li:1999:CSFW}; or combine policies
through special operators which decide on the policies'
applicability \cite{ribeiro99a}.

These mechanisms are used to solve conflicts resulting from the
existence of implicit rules in common language. For instance, a
user specifies that all his files should not be read by any one
else, and simultaneously, he specifies that the files with
information about a particular project should be accessible by all
members of the project. This situation is not a conflict in common
language since the second rule is obviously an exception to the
first, but it may be a conflict within a formal security
specification.

However, these conflict solving mechanisms should not be used to
solve real inconsistencies derived from unification of several
policies from several sources. In fact, they can even be
detrimental, because they can masquerade real inconsistencies and
produce wrong results.

Although conflicts between contradictory policies are the most
important type of inconsistency that may be present in a global
security policy, they are not the only ones. For instance, a
policy may be completely overridden by another policy in such a
way that the former policy is completely useless; or the
combination of two or more policies may result in a policy that
denies every action in the system.

Furthermore, within an organization, it is not enough to verify
the security policy self-consistency, it is also necessary to
verify the consistency of the security policy with other
specifications of the organization. For instance, if an
organization's workflow application requires access to some
documents and the security policy forbids that access, then the
security policy is inconsistent with that workflow specification,
which may prevent the organization from working as expected.

In fact, given the constraint nature of security policies, any
specification document of an organization which comprises one or
more constraints, may be a source of inconsistencies. Thus, the
search for security policy inconsistencies does not have a closed
solution valid for every organization and situation. Each
organization may have different specifications with constraints
and different interpretations of security policy inconsistencies.

This paper's contribution is twofold: First, we address the
general problem of checking for security policy inconsistencies,
whatever they are, on large complex policies; then we address the
problem of finding inconsistencies between security policies and
other specifications with constraints, namely workflow
specifications. Both problems are addressed within a novel
approach comprising a tool developed by us (PCV -- Policy
Consistency Verifier), based on the CHR constraint language
\cite{Fru98}, which finds inconsistencies within and between
security policies and other specifications.

The rest of the paper is organized as follows. We first briefly
describe the CHR language. Then describe the tool architecture.
Sections \ref{sec:secch} and \ref{sec:wfch} describe how security
policy and workflow specifications are handled by the tool.
Finally, in section \ref{sec:related} we briefly survey some
related work, and in section \ref{sec:conclusion} we conclude the
paper.

\section{Constraint Handling Rules}
\label{sec:chrs}

CHR is a high-level language designed for writing user\--defined
constraint systems \cite{Fru98}. CHR is essentially a
committed-choice language consisting of guarded rules that
rewrite constraints into simpler ones until they are solved. CHR
rules are of two types: simplification rules and propagation
rules. Simplification rules replace user-defined constraints by
simpler ones. Propagation rules add new redundant constraints
that may be necessary to do further simplifications.

\mathfig{fig:chr}{Example of CHR rules.} { $
\underbrace{\overbrace{A \leq B}^{Constraint}, B \geq A}_{Head}
\Leftrightarrow \underbrace{true}_{Guard} \; | \;
\underbrace{A = B}_{Body}$ //Simplification rule \\
\\
$A \leq B ,  B \leq C \Rightarrow  true \; | \; A \leq C$
// Propagation rule \\
\\
$X<Y \setminus X\neq Y \Leftrightarrow true.$ // Simpagation rule
 }

A CHR rule consists of three parts: a head, a guard and a body
(Figure \ref{fig:chr}). Each part is a conjunction of constraints.
There are two kinds of constraints, user-defined and built-in
constraints. User-defined constraints are relations that must
hold between one or more entities, which assume the form of
predicates or operations over those entities (e.g. less(1,2) or 1
$<$ 2). Built-in constraints are simple constraints that can be
directly solved by the underlying solver (e.g. A = B).

A simplification rule works by replacing the constraints in the
rule's head by the constraints in the rule's body, provided the
constraints in the rule's guard are true. A propagation rule adds
the constraints in the body but keeps the constraints in the head
(Figure \ref{fig:chr}). Figure \ref{fig:chr} shows also a third
type of rule named ``simpagation". A ``simpagation" rule is
equivalent to a simplification rule with some of the heads
repeated in the body. On a ``simpagation" rule only the heads
after the ``$\setminus$" sign are removed.

\section{Overview of PCV}
\label{sec:architecture}

Security policies can be seen as collections of constraints more
or less structured (depending on the security platform used) into
complex rules and policies. These constraints may be as simple as
``A middle manager cannot approve purchases over a specified
amount", or they can be as complex as constraints comprising
forms of prohibition, permission or obligation of user actions.

Given the constraint nature of security policies, the PCV
verifier is a natural candidate to be implemented with a
constraint language such as the CHR language. This approach
simplifies tool construction and potentiates its extensibility to
other inconsistency definitions.

\begin{figure}[htb]
  \centering
  \includegraphics[width=\columnwidth]{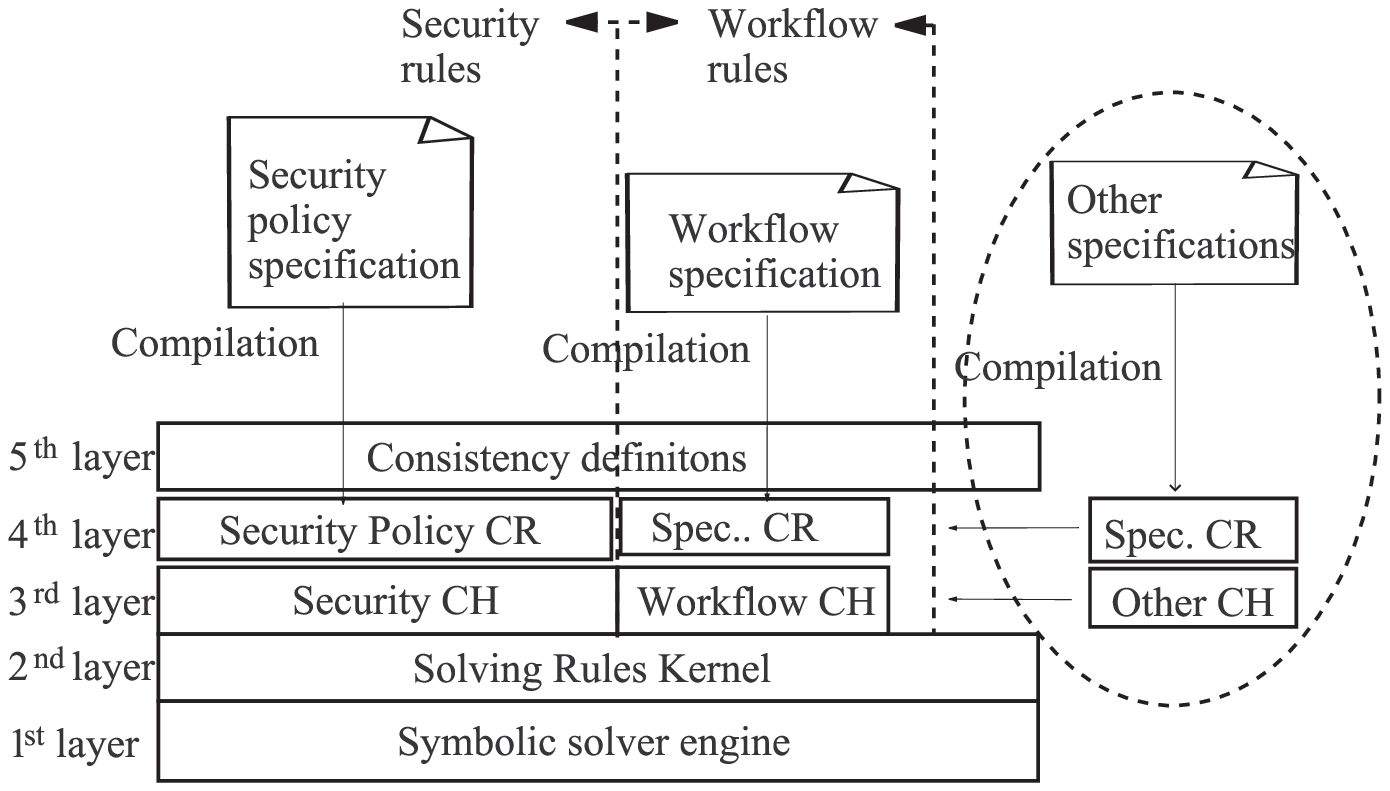}
  \caption{The consistency verifier architecture.
           CH~stands for Constraint Handler, and CR~stands for Constraint Rules}
  \label{fig:architecture}
\end{figure}

PCV is composed by five layers (Figure \ref{fig:architecture}).
The first layer is the CHR symbolic solver engine, which is the
only layer not comprised of CHR rules. The second layer is
composed by the rules which handle basic constraint predicates
(e.g. $A \leq B$, $a \in G$) and constitutes the verifier's
kernel. The third layer contains the rules which comprise the
knowledge on how to decompose the specific constraints placed by
each type of specification (security, workflow), into basic
constraints. The fourth layer contains rules resulting from the
compilation of the different specifications (e.g. security policy
specification, workflow specification). Finally, the fifth layer
contains the verifier goals, with the definitions of the security
inconsistencies being searched.

The purpose of this layering approach is threefold: (i) it
simplifies the handler design, because each type of constraints
can be handled independently; (ii) it simplifies the proof of
correctness, because each layer has no knowledge of the layers on
top and see the constraints of lower layers as built-ins; (iii)
it simplifies the addition of new specification handlers, by
defining the rules required by each specification.

Briefly, the process by which inconsistencies are found works as
follows. First the constraints within each specification being
verified are compiled into CHR rules. Second, the PCV verifier is
invoked with the constraint goals comprising the inconsistency
definitions. These goals are successively decomposed into simpler
constraints, by the rules generated by the compilers (fourth
layer), and by the rules comprising the knowledge on each
specification type (third layer), until they can be solved by the
kernel rules.

In the following sections we proceed by describing each of these
layers. For the sake of simplicity, each of the third, fourth and
fifth layers were split in two (Figure \ref{fig:architecture}),
separating the rules of each layer that handle security
constraints from the rules that handle other constraints.

\section{Kernel rules} \label{sec:basics}

The verifier's kernel rules are responsible for handling basic
constraints, resulting from the application of simple operators to
basic entities.

Kernel rules are divided into two major groups: rules to handle
order and equality operators ($>$, $<$, $\leq$, $\geq$, $=$,
$\neq$); and rules to handle membership and set constraints
($\in$, $\notin$, $\cup$ (union), $\cap$ (meet), $\#$
(cardinality), $[\;]$ (index)).

\subsection{Order and equality rules}

The rules to handle order and equality constraints derive from
the ``minmax" handler proposed by \cite{Fru98}, augmented with
optional temporal qualifiers but without the ``min" and ``max"
constraints.

A constraint may be timed or timeless. A timed constraint results
from applying a temporal qualifier to a timeless constraint (e.g.
$X < Y$ becomes $X < Y \at T$). Timed constraints, by opposition
to timeless constraints, are only valid at an instant T.

Knowing how to handle timed constraints is of utmost importance
for checking for security policy inconsistencies. Many security
policies use the notion of time when specifying dependencies from
past and future events. For instance, history-based policies like
the Chinese Wall policy \cite{ChineseWall} use the notion of time
to allow access to an object only if the same user has not
accessed another object in the same class of interest. Another
example is given by obligation-based policies which use the
notion of time to ensure that some action happens in the future
\cite{ribeiro2000b}.

\begin{figure}[htb]
\begin{center}\small
\[H_1,\ldots,H_n \Rightarrow G | B_1,\ldots,B_m.\]
derives
\[^{2^{n}-1} \left\{ \setlength\arraycolsep{1pt}
\begin{array}{lcl}
H_1 \at T,\ldots,H_n & \Rightarrow & G | B_1 \at T,\ldots,B_m \at T. \\
&\ldots \\
H_1,\ldots,H_n \at T & \Rightarrow & G | B_1 \at T,\ldots,B_m \at T. \\
&\ldots \\
H_1 \at T,\ldots,H_n \at T & \Rightarrow & G | B_1 \at
T,\ldots,B_m \at T.
\end{array}
\right.
\]
and
\[H'_1,\ldots,H'_n \Leftrightarrow G' | B'_1,\ldots,B'_m.\]
derives
\[^{2^{n}-1} \left\{ \setlength\arraycolsep{1pt}
\begin{array}{lcl}
H'_n,\ldots\setminus H'_1 \at T & \Leftrightarrow & G' | B'_1 \at T,\ldots,B'_m \at T. \\
&\ldots \\
H'_1,\ldots\setminus H'_n  \at T & \Leftrightarrow & G | B'_1 \at T,\ldots,B'_m \at T. \\
&\ldots \\
H'_1 \at T,\ldots,H'_n \at T & \Leftrightarrow & G | B'_1 \at
T,\ldots,B'_m \at T.
\end{array}
\right.
\]
\end{center}
\caption{Template rules to build timed propagation and timed
simplification rules from their timeless counterparts.}
\label{fig:time}
\end{figure}

The rules which handle timed constraints derive from the rules
which handle timeless constraints in accordance with the templates
in Figure \ref{fig:time}. A timeless rule with $n$ constraints in
its head derive $2^n-1$ timed rules, each with a different
combination of timed and timeless heads. The template rules for
timed simplification rules are slightly different from the
template rules for timed propagation rules. Timed simplification
rules only remove timed constraints. The timeless constraints
used to activate each rule are not removed to preserve the
activation sequence of timeless rules and therefore maintain their
correctness.

\mathfig{fig:equal}{Rules to handle timed equality. $(\at T)$
stands for an optional time qualification. Although strict CHR
rules do not allow optional elements, they are used here to
simplify the description.}{ commutativity \=@ \kill

built\_in \>@ $\eqt{X}{Y} \Leftrightarrow ground(X),ground(Y) | X=Y.$\\
reflexivity  \>@ $\eqt{X}{X}  \Leftrightarrow true.$\\
commutativity \>@ $\eqt{X}{Y} \setminus \eqt{Y}{X} \Leftrightarrow true.$ \\
subsumption \>@ $\eqt{X}{Y} \setminus \leqt{Y}{X} \Leftrightarrow \neqii{X}{Y} | true.$ \\
subsumption \>@ $\eqt{X}{Y} \setminus \leqt{X}{Y} \Leftrightarrow \neqii{X}{Y} | true.$ \\
irreflexivity \>@ $\eqt{X}{Y}, \lssot{Y}{X} \Leftrightarrow fail.$ \\
irreflexivity \>@ $\eqt{X}{Y}, \lssot{X}{Y} \Leftrightarrow fail.$ \\
tautology \>@ $\eqt{X}{Y}, \neqot{Y}{X} \Leftrightarrow fail.$ \\
tautology \>@ $\eqt{X}{Y}, \neqot{X}{Y} \Leftrightarrow fail.$ \\
\\
\% Transitivity rules\\
WithSelf \>@ $\eqt{X}{Y}, \eqt{X}{Z} \Rightarrow  \neqiii{X}{Y}{Z}  | \eqt{Y}{Z}.$ \\
WithSelf \>@ $\eqt{X}{Y}, \eqt{Y}{Z} \Rightarrow  \neqiii{X}{Y}{Z}  | \eqt{X}{Z}.$ \\
WithSelf \>@ $\eqt{X}{Y}, \eqt{Z}{X} \Rightarrow  \neqiii{X}{Y}{Z}  | \eqt{Y}{Z}.$ \\
WithSelf \>@ $\eqt{X}{Y}, \eqt{Z}{Y} \Rightarrow  \neqiii{X}{Y}{Z}  | \eqt{X}{Z}.$ \\
WLessOrEqual \>@ $\eqt{X}{Y}, \leqot{X}{Z} \Rightarrow \neqiii{X}{Y}{Z} | \leqt{Y}{Z}.$ \\
WLessOrEqual \>@ $\eqt{X}{Y}, \leqot{Y}{Z} \Rightarrow \neqiii{X}{Y}{Z} | \leqt{X}{Z}.$ \\
WLessOrEqual \>@ $\eqt{X}{Y}, \leqot{Z}{X} \Rightarrow \neqiii{X}{Y}{Z} | \leqt{Z}{Y}.$ \\
WLessOrEqual \>@ $\eqt{X}{Y}, \leqot{Z}{Y} \Rightarrow \neqiii{X}{Y}{Z} | \leqt{Z}{X}.$ \\
\>$\vdots$
 }

Although the rules generated by the application of the template
rules of Figure \ref{fig:time} are sufficient to handle every
timed constraint derived from a user-defined timeless constraint,
they cannot handle the timed constraints derived from built-in
timeless constraints. While timeless-equality constraints are
handled by the underlying built-in solver, the same is not true
for timed-equality constraints ($\eqt{X}{Y}$) which require rules
such as the ones in Figure \ref{fig:equal}. These rules can be
divided in two groups: the first is composed of simplification
rules describing redundancies and conflicts between timed equality
constraints and other constraints; the second consists of
propagation rules describing the transitivity properties between
timed equality constraints and other constraints.

\subsection{Set and membership rules}

\mathfig{fig:meet}{Basic rules for membership and meet
constraints.}{ commutativity \=@ $C=A\cap B \setminus C=B\cap A
\Leftrightarrow true.$ \kill
tautology \>@ $X \in G, X \notin G \Rightarrow fail.$ \\ \\
identity \>@ $C=A\cap A \Leftrightarrow C = A.$ \\
commutativity \>@ $C=A\cap B \setminus C=B\cap A \Leftrightarrow
true.$
\\\\
distributivity \>@ $X\in C, C=A\cap B \Rightarrow A\neq B | X\in A,\: X\in B.$ \\
revDist \>@ $X\in A, X\in B, C=A\cap B \Rightarrow A \neq B | X\in C.$\\
revNotDist \>@ $X\notin A, C=A\cap B \Rightarrow X\notin C.$ \\
revNotDist \>@ $X\notin B, C=A\cap B \Rightarrow X \notin C.$\\
notDistrib \>@ $X\notin C, X\in A, C=A\cap B \Rightarrow X \notin B.$\\
notDistrib \>@ $X\notin C, X\in B, C=A\cap B \Rightarrow X \notin
A.$ \\ \\
revDist \>@ $labeling, $\=$X\in A, C=A\cap B \Rightarrow A\neq B |$ \\
\>\>$((X\notin B, X\notin C) ; (X\in C, X\in B)).$ \\
revDist \>@ $labeling, X\in B, C=A\cap B \Rightarrow A\neq B
|$ \\
\>\>$( (X\notin A, X\notin C) ; (X\in A, X\in C) ).$ \\
notDistrib \>@ $labeling, X\notin C, C=A\cap B \Rightarrow A \neq
B |$ \\
\>\>$(X\notin A; X\notin B).$ }

\emph{Set} constraints are not handled as usual. Since the
verifier is to be used primarily on non-instantiated
specifications, when most sets are yet undefined --in the sense
that their members are not yet known-- it is not possible to
directly solve in order to their contents. Instead of solving
\emph{set} constraints, we use them to derive \emph{membership}
constraints which can be directly solved. For instance, using the
``distributivity" rule, followed by the ``tautology" rule of
Figure \ref{fig:meet}, the goal $\state{C=A\cap B, X \in C, X
\notin B}$ leads to a \emph{fail} state.

\emph{Membership} constraints may also be time-qualified such as
order constraints. The CHR rules to handle such constraints are
also derived from the template rules of Figure \ref{fig:time} but
applied to the CHR rules which handle timeless \emph{membership}
constraints. Timed \emph{membership} constraints are very useful
when sets' contents are dynamic, which happens to be frequent in
security policies, e.g. a user may have been playing a role and
now he is playing another. On the other hand we do not provide
rules to handle timed \emph{set} constraints, since most
relations between sets used in security policies (e.g. $C=A\cap
B$) are fixed during the whole policy lifetime.

The last three rules of figure \ref{fig:meet} contain
disjunctions in their bodies (the ';' stands for built-in
disjunction), which must be handled by test and backtracking. In
order to improve efficiency these rules should be delayed until
there are no more constraints in the goal to simplify. The special
$labeling$ constraint is the last constraint in the goal to be
solved. Thus, using this constraint in the head of rules ensures
that they are activated only when all other constraints have been
already simplified.

Without further assumptions the program composed by the rules in
Figure \ref{fig:meet} does not terminate, because the constraints
generated by propagation rules, may be used to generate other
constraints which are going to enable again the same rules. For
instance, the constraints derived by the ``distributivity" rule
could be used by the ``revDist" rule to derive the constraint
$X\in C$, which may be used again to fire the ``distributivity"
rule.

However, it is possible to ensure the program termination under
three new assumptions:
\begin{itemize}
  \item The CHR solver verifies the constraint store, before introducing
  new constraints, to prevent the existence of multiple copies of
  the same constraint in the constraint store\footnote{Most CHR
  solvers can be instructed to perform this check by enabling the
  ``already\_in\_store" option.}.
  \item Membership constraints are never removed from the
  constraint store.
  \item Meet constraints cannot be added to the constraint store,
  by any program's rule.
\end{itemize}

Under the first assumption the number of \emph{membership}
constraints in the goal store is bounded, provided that the
number of variables in the initial goal store is bounded, and
that none of the rules derives constraints with new variables.
The second assumption is necessary to ensure monotonicity of
\emph{membership} constraints in the constraint store. The third
assumption is necessary to ensure monotonicity and boundedness of
\emph{meet} constraints in the constraint store.

\mathfig{fig:restriction}{Rules for set restriction constraints.
The constraint $C=A:R$ is a short form for $C=\{x \in A : R(x)\}$
where $C$ and $A$ are sets and $R$ is a boolean function over
$x$}{ idempotent \=@ \kill
idempotent \>@ $C=A:R(\_) \setminus D=A:R(\_) \Leftrightarrow C=D.$ \\ \\
restriction \>@ $X\in C, C=A:R(_) \Rightarrow R(X), X\in A.$ \\
revRestric \>@ \=$labeling, X\in A, C=A:R(_) \Rightarrow$ \\
\>\>$(( R(X), X \in C);(X \notin C, \neg R(X))).$ }

The rules in Figure \ref{fig:restriction} are used to solve
constraints using the restriction operand($:$). The restriction
operand is a binary operand between a set and a boolean function.
The operation defines a new set comprised by all the members of
the first set which satisfy the boolean function. The strategy
followed by these rules is the same as the one followed by the
rules to handle \emph{meet} constraints (Figure \ref{fig:meet}).
The \emph{restriction} constraints are used together with
\emph{membership} constraints to derive other \emph{membership}
constraints, and the rules containing disjunctions are delayed
until the end of the simplification process.

\mathfig{fig:cardinality}{Rules to handle cardinality
constraints. The cardinality of set $S$ is denoted by $|S|$.}{
\% Relation with other set constraints \\
eqSetMin \=@ \kill
identity \>@ $N1=|L| \setminus N2=|L| \Rightarrow N1=N2.$ \\
meet \>@ $NC=|C|, NA=|A|, C=A\cap B \Rightarrow NC \leq NA.$ \\
meet \>@ $NC=|C|, NB=|B|, C=A\cap B \Rightarrow NC \leq NB.$ \\
join \>@ $NC=|C|, NA=|A|, C=A\cup B \Rightarrow NA \leq NC.$ \\
join \>@ $NC=|C|, NB=|B|, C=A\cup B \Rightarrow NB \leq NC.$ \\
restrict \>@ $NA=|A|, NC=|C|, C=A:R \Rightarrow NC \leq NA.$ \\
less \> $N=|A| \Rightarrow integer(N), N<0 |\; fail.$ \\
\\
\% For defined sets. Sets with a specific length.\\
eqSetMin \>@ $N=|L| \Rightarrow is\_list(L) |\; length(L,N).$ \\
\\
\% For undefined sets. Sets without a specific length.\\
insert \>@ $X \in L, N=|L| \Rightarrow \neg is\_list(L) |\;
member(X,L).$ \\
eqSetMin \>@ $labeling, $\=$N=|L| \Rightarrow$ \\
\>\>$\neg is\_list(L), integer(N), N \geq 0 |$ \\
\>\>$cardinal(L,N).$ \\
lesser \>@ $labeling, N=|L|, N < N1 \Rightarrow$ \\
\>\>$\neg is\_list(L),integer(N1), N1 > 0 |$ \\
\>\>$cardinal(L,N1).$ \\
lesseq \>@ $labeling, N=|L|, N \leq N1 \Rightarrow$ \\
\>\>$\neg is\_list(L),integer(N1), N1 \geq 0 |$ \\
\>\> $cardinal(L,N1).$ }

The rules to handle cardinality constraints (Figure
\ref{fig:cardinality}) may be divided into three groups. The first
group is used to derive inequality constraints between
cardinality values of sets related by \emph{meet}, \emph{union}
and \emph{restriction} constraints. The second group is composed
of only one rule and is used to translate a cardinality constraint
on a defined set into the built-in predicate $length$. This rule
is used only if the set is completely defined, in the sense that
all its members are known. However, as explained before, most sets
are not completely defined at verification time, which makes it
impossible to know their cardinality. The third group of rules is
used to verify if at the end of the verification process, the
number of elements known to be in a set, over which a upper bound
cardinality constraint exists does not exceed that upper bound.

\section{Security Rules} \label{sec:secch}

As described in section \ref{sec:architecture} the rules which
handle specific security constraints are divided into three
related layers: the security constraint handler; the security
policy rules resulting from the compilation of the security
specification; and the consistency definition goals. Each of
these layers is dependent on the preceding and following layer
and all are dependent on the specificities of the security policy
definition environment. The security policy definition
environment used in the current prototype is the Security Policy
Language (SPL) \cite{ribeiro99a}. This security language was
developed by us with the purpose of specifying security controls
for complex environments where several security policies must be
enabled simultaneously.

In the remaining of this section we briefly describe SPL. Then we
describe how the security CH handles the types of constraints
placed by SPL. Finally, we describe the process of compiling SPL
to CHR using the rules provided by the security CH.

\subsection{SPL} \label{sec:SPL}

SPL is a security language designed to express policies to decide
about events' acceptability. An event's acceptability depends on
the properties of the event (e.g. author, target and action), on
the context at that time and on the properties of past and future
events. SPL entities are typed entities with an explicit
interface by which their properties can be queried. Some of the
entities manipulated by SPL are internal, such as rules and
policies, but most are external, like users, files, actions,
objects and events. The properties of each external entity
depends heavily on the platform (operating system, workflow
engine) implementing it.

The language is organized in a hierarchical delegation tree of
security policies, in which the master policy is the root
delegation starting point. A SPL policy is a structure composed by
sets and rules, whose purpose is to express simple concepts like
``separation of duty'',  ``information flow'', or  ``general
access control''.

Sets contain the entities used by the policies to decide on event
acceptability. A SPL rule is a function of events that can assume
three values:  ``allow'',  ``deny'' and  ``notapply''. It's
purpose is to decide on the acceptability of the current event. A
rule can be simple or composed. A simple SPL rule is a tuple of
two logical expressions. The first logical expression decides on
the applicability of the rule, and the second decides on the
acceptability of the event.

A SPL rule can be composed by other SPL rules through a specific
tri-value algebra with five logic operations: conjunction (AND);
disjunction (OR); negation (NOT); existential quantification
(EXIST $x$ IN $set$ {$rule$}); and universal quantification
(FORALL $x$ IN $set$ {$rule$}).  These operations behave similarly
to their binary homonyms if the  ``notapply'' value is not used
and the  ``allow'' and  ``deny'' values are used as true and
false, respectively.

Each policy has one special SPL rule called the  ``query rule''
which is identified by a query mark before the name that defines
the policy behavior.

\mathfig{fig:private}{Simple policy stating that objects belonging
to IDocs can only be sent to users belonging to OrgUsers.} {
\hspace{1ex} \= event.action = ``SendEmail" \& \=    //
Applicability \kill
policy Private( user set OrgUsers ) \{ \\
\> object set IDocs; \>     // Policy data \\
?Private:   \>\>                   // Rule name. \\
\> event.action = ``SendEmail" \& \>    // Applicability \\
\> event.target IN IDocs \>    // expression. \\
\>   ::                      \>    // Separation marker. \\
\> event.par[1] IN OrgUsers \> // Aceptability \\
\}                   \> \>            // expression. }

Figure \ref{fig:private} shows a simple policy stating that
documents internal to the entity defining the policy cannot be
sent to someone outside the organization. The policy has two sets
and one SPL rule: the query rule. One of the sets is a policy
parameter and contains the users that belong to the organization.
The other is internal to the policy and contains the department's
internal documents. The rule uses the special variable ``ce'' to
access the current event properties. The rule's applicability
expression states that the policy is defined only for events
whose targets are department's internal documents internal and
whose action is to send an e-mail. The rule's acceptability
expression states that for those events that satisfy the
applicability expression the only allowed events are the ones
that send the e-mail to a user of the organization.

\subsection{Security Constraint Handler}\label{sec:splrules}

Although most SPL constraints can be handled directly by the
verifier's kernel rules, some cannot. For instance, the
constraints resulting from the logical negation of other
constraints, or the constraints resulting from using the SPL
tri-logical operators over SPL rules, require some additional
rules in order to be solved.

\mathfig{fig:logical}{Rules to handle logical constraints. The
symbol $\xor$ stands for exclusive disjunction.}{
comutativity \=@ $A \wedge B \setminus B \wedge A \Leftrightarrow true.$ \\
comutativity \>@ $A \vee B \setminus B \vee A \Leftrightarrow true.$ \\
comutativity \>@ $A \xor B \setminus B \xor A \Leftrightarrow true.$ \\
\\
definition \>@ $labeling \setminus A \vee B \Leftrightarrow \neqii{A}{B} | (A ; B).$ \\
definition \>@ $A \wedge B \Leftrightarrow \neqii{A}{B} | A , B.$ \\
definition \>@ $A \xor B \Leftrightarrow \neqii{A}{B} |(A \wedge \neg B) \vee (\neg A \wedge B).$ \\

identity \>@ $A \wedge A \Leftrightarrow A.$ \\
identity \>@ $A \vee A \Leftrightarrow A.$ \\
irreflexivity \>@ $A \xor A \Leftrightarrow fail.$ \\

tautology \>@ $\neg true \Leftrightarrow fail.$ \\
tautology \>@ $\neg fail \Leftrightarrow true.$ \\
tautology \>@ $\neg (\neg A) \Leftrightarrow A.$ \\
deMorgan \>@ $\neg (A \wedge B) \Leftrightarrow (neg A \vee neg B).$ \\
deMorgan \>@ $\neg (A \vee B) \Leftrightarrow \neg A,  \neg B.$ \\
definiton \>@ $\neg (A \xor B) \Leftrightarrow (A \wedge B) \vee (\neg A \wedge \neg B).$ \\
\\
reduction \>@ $\neg (A < B (\at T)) \Leftrightarrow B \leq A (\at T).$ \\
\>$\vdots$ }

Handling logical negation is straightforward, provided that the
verifier also handles logical constraint conjunction and
disjunction. The rules which handle these constraints are shown
in Figure \ref{fig:logical}. To simplify SPL to CHR compilation
we also provide some rules to handle exclusive disjunction over
constraints.

Although logical constraint conjunction and disjunction are
handled by direct translation to their built-in counterparts (by
the ``definition" rules), their negations are handled by the
DeMorgan rules, thus pushing negations to basic kernel
constraints where they can be handled by the ``reduction" rules.

\mathfig{fig:spland}{Rules to handle SPL's tri-logical conjunction
and negation.}{ definition \=@ \kill
definition \>@ $R = \notr r(D,A) \leftrightarrow R=r(D,\neg A).$ \\
\\
commutativity @ $R3=R1 \andr R2 \setminus R4=R2 \andr R1 \Leftrightarrow R4=R3.$ \\
identity \>@ $R3 = R1 \andr R1 \Leftrightarrow R3=R1.$ \\
neutral   \>@ $R3 = r(fail,X) \andr R2 \Leftrightarrow R3=R2.$ \\
neutral   \>@ $R3 = R1 \andr r(fail,X) \Leftrightarrow R3=R1.$ \\
absorb    \>@ $R3 = r(true,fail) \andr R2 \Leftrightarrow R3=r(true,fail).$ \\
absorb    \>@ $R3 = R1 \andr r(true,fail) \Leftrightarrow R3=r(true,fail).$ \\

default \>@ $R3 = r(true,true)$\=$\andr r(D2,A2) \Leftrightarrow$ \\
\>\> $R3 = r(true, \neg D2 \vee A2).$ \\
default \>@ $R3 = r(D1,A1)$\=$\andr r(true,true) \Leftrightarrow$ \\
\>\> $R3 = r(true, \neg D1 \vee A1).$ \\
definition \>@ \=$R3 = r(D1,A1) \andr r(D2,A2) \Leftrightarrow$ \\
\>\> $R3 = r(D1 \vee D2, (\neg D1 \vee A1) \wedge (\neg D2 \vee A2)).$ }

The rules that handle constraints resulting from applying SPL
tri-logical operators to SPL rules are straightforward. Most of
these rules are just translations of each operator's definition
(Figure \ref{fig:spland}). For instance, the tri-logical
conjunction of two SPL rules defined by the predicates $r(D1,A1)$
and $r(D2,A2)$, in which $D1$ and $D2$ stand for the domain
expressions of each of the rules and $A1$ and $A2$ stand for the
acceptability expressions, is simplified to another SPL rule
defined by the predicate $r(D1 \vee D2, (\neg D1 \vee A1) \wedge
(\neg D2 \vee A2))$ (definition rule in Figure \ref{fig:spland}).

The remaining rules reflect special situations in which the
result is known without the need to evaluate the definition. For
instance, the result of a tri-logical conjunction between two SPL
rules in which one of them has an empty domain is equal to the
other rule ($R \andr r(fail,A)$ is $R$).

\mathfig{fig:splforall}{Rules to handle SPL's tri-logical
universal operator.}{ convert \=@ \=f\kill
\% Quantification over proper sets \\
empty \>@ $\forallr(Set,TR,R) \at T \Leftrightarrow is\_list(Set), Set=[\:] |$ \\
\>\>$R = r(fail,true).$ \\
forEach \>@ $\forallr(Set,TR,R) \at T \Leftrightarrow is\_list(Set),$ \\
\>\>$Set=[X|Tail] | R = \call TR(X), R = R1 \andr R2,$ \\
\>\>$\forallr(Tail,TR,R2) \at T.$ \\
\\
\% Quantification over undefined sets \\
convert \>@ $\forallr(Set,Tr,R) \at T \Leftrightarrow \neg is\_list(Set) |$ \\
\>\>$\forallr(Set,Tr,R,[]) \at T.$ \\
insert \>@ $X \in Set \at T \setminus \forallr(Set,Tr,R,U) \at T\Leftrightarrow$ \\
\>\>$not\_member(X,U) | R = Tr(X) \andr R2,$ \\
\>\>$\forallr(Set,Tr,R2,[X|U]) \at T.$ \\
insert \>@ $X \in Set \setminus \forallr(Set,Tr,R,U) \at T \Leftrightarrow$ \\
\>\>$not\_member(X,U) | R = Tr(X) \andr R2,$ \\
\>\>$\forallr(Set,Tr,R2,[X|U]) \at T.$ \\
\\
no\_more \>@ $labeling \setminus \forallr(Set,Tr,R,U) \at T \Leftrightarrow$ \\
\>\>$R = r(fail,true).$  }

SPL tri-logical quantifiers are slightly more complex to handle.
Figure \ref{fig:splforall} shows the rules to handle the universal
quantifier $\forallr(Set,Tr,R) \at T$, which should be read as $R
= \{\forall_x \in Set \at T : Tr(x)\}$. The rules are divided into
two sets: the rules that handle universal quantifiers over
defined sets; and the rules that handle universal quantifiers over
sets defined by membership constraints.

Both sets of rules handle the quantifiers constraints by unfolding
them to $n$ tri-logical conjunctions. However, the rules that
handle quantifiers over undefined sets require an extra constraint
property to account for the membership constraints which have
already been used with each quantification constraint (Figure
\ref{fig:splforall}). This account is necessary to prevent
non-termination caused by using each membership constraint more
than once with each quantification constraint.

The rules to handle existential quantifier constraints are very
similar to the ones that handle universal quantifier constraints.
The difference is that existential quantifiers are unfolded to
tri-logical disjunctions.

\subsection{Compiling SPL}

For the purpose of consistency verification, SPL policies should
be seen as operator definitions, which are used to state event
constraints. The goal of the SPL compiler is to generate the
rules necessary to handle each of these types of constraints.

Since SPL is a constraint language, translating it into CHR is a
direct process. Each policy is seen as a complex constraint
composed by other simpler constraints. Thus, each policy
definition is translated into one simplification rule, with one
user-defined constraint in the head, several constraints in the
body and no guard. The constraint in the head is a predicate with
the policy's name, applied to a tuple with a SPL rule
representing the policy, every explicit policy's parameters and
four implicit ones: current event, global and local variables.
The body of the rule is composed by one constraint for each SPL
group or rule definition inside the policy. These constraints can
then be further simplified by the rule of the consistency engine.

 \mathfig{fig:chrprivate}{The figure shows the
translation of the SPL policy of Figure \ref{fig:private} to
CHR.} {
private(\=$Event$, OrgUsers, Locals, Globals, R) $\Leftrightarrow$\\
\>Locals \== \kill
\>Event \>= event(Actor,Action,Target,Pars, Time),\\
\>Locals \>= locals(private\_vars(IDocs)), \\
\>R \>= r( \=Action =  ``SendMail'' $\wedge$ Target $\in$ IDocs, \\
\>\>\> Pars[1] $\in$ OrgUsers). }

Figure \ref{fig:chrprivate} shows the translation of the SPL
policy presented in figure \ref{fig:private}. The policy is
translated into a simplification rule with two constraints in the
body. One of the constraints states that the set ``IDocs'' is
defined inside the predicate ``locals'', in the ``privat\_vars''
section. The other states that the SPL rule to apply is defined
by the predicate ``r'' applied to the tuple composed by the
applicability and acceptability expressions.

Although most policies can be translated into a single CHR rule,
some require more than one rule, and some require special
handling to increase performance. For instance, as referred in
the previous section, existential quantifiers are unfolded to
tri-logical disjunctions of each of the quantification's elements,
which in turn derive built-in disjunctions that may compromise
performance due to backtracking. In some situations, existential
quantifiers are transformed into conjunctions of two constraints
by a process known as Skolemization \cite{hooger90}. This
transformation is possible if the set of the quantification is
not empty and the applicability expression of the SPL rule does
not depend on the quantification variable. Under these conditions,
the existential quantification
 ``$\exists_x \in A : rule(x)$'' is equivalent to  ``$c \in A$ and
$rule(c)$'' where  ``$c$'' is a Skolem constant.

\subsection{Security self-consistency} \label{sec:self-defs}

A security policy may be inconsistent in several ways. For
instance, a security policy which is never applicable is
unnecessary, thus inconsistent. On the other hand, a security
policy that denies every event is also inconsistent. Several
other inconsistency definitions may be devised. Currently, our
prototype is able to check four types of policy inconsistencies:

\begin{itemize}
  \item Inapplicability: the policy is never applicable;
  \item Monotonic denial: the policy denies every event;
  \item Monotonic acceptance: the policy accepts every event;
  \item Rule redundancy: one or more rules in the policy are
  redundant.
\end{itemize}

Verifying the first three types of inconsistency is
straightforward. It is only necessary to find a solution for the
$event$ variable on each of the following goals:
{\small
\begin{align*}
&E \in AllEvents,\; myPolicy(E,\ldots,r(D,A)),\; D.\\
&E \in AllEvents,\; myPolicy(E,\ldots,r(D,A)),\; \neg D \vee A.\\
&E \in AllEvents,\; myPolicy(E,\ldots,r(D,A)),\; \neg D \vee \neg A.
\end{align*}
}

The fourth type is slightly more complex. Briefly, the verifier
should replace, in turn, each of the rules to check for
redundancy, by a dummy rule with an empty applicability domain,
and check for differences between the original policy and the
modified policy.

\mathfig{fig:diff}{Rules to handle the $\diff$ operator. This
operator restraints two SPL rules to be different}{ commutativity
\=@ \kill
commutativity \>@ $R1 \diff R2 \ R2 \diff R1 \Leftrightarrow true.$ \\
identity \>@ $R \diff R \Leftrightarrow fail.$ \\
definition \>@ \=$r(D1,A1) \diff r(D2,A2) \Leftrightarrow$ \\
\>\> $(D1 \xor D2) \vee ((D1 \wedge A1) \xor (D2 \wedge A2)).$ }

The replacement of each rule by the dummy rule is done by the
underlying platform. The actual test for policy differences is
done by the $\diff$ operator (Figure \ref{fig:diff}), which
restraints the ``query" rules of each policy to be different. If
this constraint fails, this means that both the original and the
modified policies are equal and the replaced rule is redundant.

\section{Workflow Rules}\label{sec:wfch}

Most specifications comprising constraints within an organization
are eligible to be checked for consistency together with the
organization security policy. The specific importance of workflow
specifications results from being usually used all over the
organization, potentially crossing different security management
domains, thus increasing the probability of occurring
inconsistencies.

Although workflow specifications are usually created by a graphic
tool, and kept inside a workflow framework in an internal format,
several workflow frameworks also support the ``Workflow Process
Description Language" (WPDL) \cite{wfml} for specification
interchange purposes. Given the interchange purpose of WPDL we
have found it to be the ideal language to test the verifier's
ability to express and handle workflow specifications.

In this section we briefly describe WPDL. Then we describe how
WPDL is translated into CHR, and why the workflow CH for WPDL is
empty. Finally, we give some examples of cross-consistency goal
definitions.

\subsection{WPDL} \label{sec:wpdl}

WPDL's main entities are: {\em activities}, {\em participants} and
{\em transitions}. Each activity is a logical, self-contained unit
of work within the workflow definition, performed by a
participant. An activity may be atomic, a sub-flow, loop activity
or a dummy activity.

Atomic activities are the ones which are going to be activated by
events and therefore controlled by the security policy. A
sub-flow activity is just a container for a sequence of
activities. A loop activity is a special control activity
comprising a loop condition. For each loop activity there are two
special transition entities pointing from and to it. The outgoing
transition points to the loop's start activity. The incoming
transition points from the loop's end activity. Dummy activities
are used to support routing decisions within the incoming and/or
the outgoing transitions.

{\em Transitions} are comprised by three elementary properties:
the from-activity, the to-activity, and the activation condition.
Transitions execution may lead to sequential or parallel
activities execution. The information related to split and join
properties is defined within the appropriate activity.

The join property decides if the activity requires the activation
of only one or every incoming transition. The split property
decides if, after the activity is performed, every outgoing
transition is activated or if only one of them is activated. In
the latter case the activated transition is chosen from a priority
list. If the first transition cannot be activated due to its
activation condition, the next transition is chosen.

{\em Participants} are resources that may be assigned to
activities. A participant may be a person, a role, an application
(automated activity), or an organizational unit.

\begin{figure}[htb]
  \centering
  \includegraphics[width=\columnwidth]{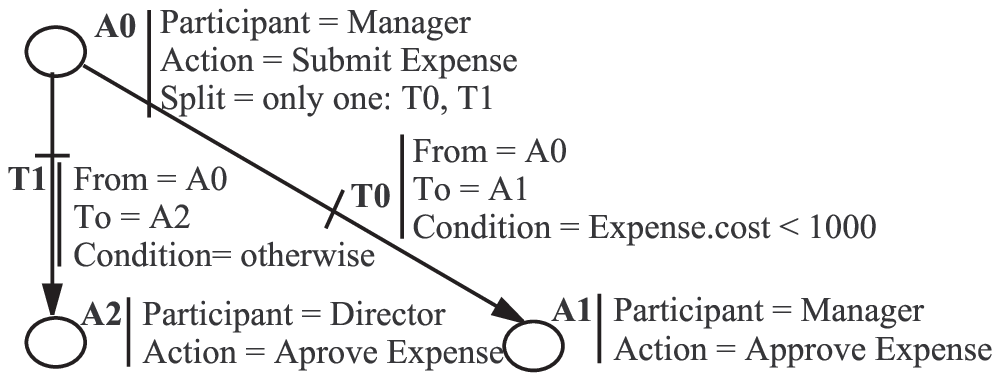}
  \caption{A workflow definition example:
  $A_i$ stands for activity $i$ and $T_i$ stands for transition $i$.}
  \label{fig:workflow}
\end{figure}

Figure \ref{fig:workflow} shows an example of a workflow
definition. ``Activity 0" is performed by a Manager and is an
expense submission. After being executed, ``activity 1" enables
two transitions, but only one of them can be executed. It first
tries ``transition 0", if the expense's cost is less than 1000
then ``transition 0" is executed and ``activity 1" becomes
enabled, otherwise ``transition 1" is executed and ``activity 2"
becomes enabled.

\subsection{Compiling WPDL}

For the purpose of consistency verification, each WPDL activity
specification is seen as the definition of a specific type of
logical operator, to state event constraints. These constraints
fail when the event is incompatible with the activity. For
instance, when the author of the event is different from the
participant required by the activity.

Transitions may also be seen as logical event operators. The
purpose of these operators is to establish temporal constraints
between events enacting the to- and the from- activities of the
transition.

The WPDL compiler generates the CHR rules for handling these
constraints. However, due to the WPDL's simplicity it is not
necessary to provide auxiliary CHR rules, such as the ones
comprising the security constraint handler (section
\ref{sec:splrules}). Thus the specific workflow constraint handler
for WPDL is empty.

\mathfig{fig:chrworkflow}{This figure shows the translation of the
workflow of Figure \ref{fig:workflow} to CHR. AllEvents, Clerk,
Budget and Boss variables are defined within Globals (not shown
for simplicity)} {
a0(\=E, Globals) $\Leftrightarrow$ \\
\>E = event(Actor,Action,Target,\_,\_), E $\in$ AllEvents,\\
\>Actor $\in$ Clerk, Action = ``Build", Target = Budget. \\
\\
a1(E, Globals) $\Leftrightarrow$ t0(E, Globals),\\
\>E = event(Actor,Action,Target,\_,\_), E $\in$ AllEvents, \\
\>Actor $\in$ Clerk, Action = ``Approve", Target = Budget. \\
\\
a2(E, Globals) $\Leftrightarrow$ t1(E,Globals),\\
\>E = event(Actor,Action,Target,\_,\_), E $\in$ AllEvents,\\
\>Actor $\in$ Boss, Action = ``Approve", Target = Budget. \\
\\
t0\_test(E, Globals) $\Leftrightarrow$ Budget = budget(Cost), Cost $<$ 1000. \\
t0(E, Globals) $\Leftrightarrow$ t0\_test(E,Globals), a0(PreviousE,Globals), \\
\>E = event(\_,\_,\_,\_, T), E $\in$ AllEvents,\\
\>PreviousE = event(\_,\_,\_,\_, PreviousT), PreviousT $<$ T. \\
\\
t1\_test(E, Globals) $\Leftrightarrow$ $\neg$ t0\_test(E, Globals). \\
t1(E, Globals) $\Leftrightarrow$ t1\_test(E, Globals), a0(PreviousE,Globals), \\
\>E = event(\_,\_,\_,\_, T), E $\in$ AllEvents,\\
\>PreviousE = event(\_,\_,\_,\_, PreviousT), PreviousT $<$ T. }

Figure \ref{fig:chrworkflow} shows the rules resulting from the
compilation of the workflow definition of Figure
\ref{fig:workflow}. Activity constraints are simplified to
conjunctions of basic constraints on event properties and of
transition constraints.

Transition constraints are simplified to basic constraints and
activity constraints. The actual constraint simplification of
transition constraints is divided into two simplification rules,
to assist in expressing dependency on other transitions, when they
are in the same split priority list. For instance, the
constraints of type $t1$ depend on the failure of the test
condition of constraints of type $t0$.

\subsection{Workflow/Security consistency} \label{sec:cross-defs}

A workflow specification may be inconsistent with a security
policy in several ways. For instance, a workflow may be
inconsistent if at least one of its activities cannot be executed
under the security policy. It may also be inconsistent if there is
no activation path between its start activity and its end activity
allowed by the security policy.

The last situation is particularly interesting since it reflects
the impossibility of performing the work for which the workflow
was conceived. To verify this inconsistency it is only necessary
to find a solution for the event variable $E$ which satisfies the
goal: {\small
\begin{multline*}
E \in AllEvents,\; lastActivity(E, \ldots), \\
\shoveright {\forallr(AllEvents, Tr(\ldots), R).} \\
\shoveleft\quad\text{where} \\
\shoveleft Tr(E,\ldots,R) \vdash masterPolicy(E,\ldots,R), close(R).
\end{multline*}
} \indent The first line of the above goal states that there is
an event, belonging to the events set, which satisfies the last
activity constraints. Since, by definition, the last activity
implies the existence of an event for each of the preceding
activities, the goal states the existence of an event for each
activity from the start activity to the end activity. The second
line of the consistency goal states that every event must also
satisfy the security policy, thus stating consistency between the
two specifications.

The ``close" constraint is an auxiliary constraint which specifies
the behaviour of the security service when a security policy does
not apply to an event. With the close assumption, the service
denies those events. With the open assumption, the service allows
those events. The rules to handle the ``close" and ``open"
constraints are defined as:
{\small
\begin{align*}
\texttt{open}\:r(D1,A1) \Leftrightarrow \neg D1 \vee (D1 \wedge A1). \\
\texttt{close}\:r(D1,A1) \Leftrightarrow D1 \wedge A1.
\end{align*}
}
\indent These rules have another important function. They act as the
bridge between the tri-value logic used by the security
constraints and the binary logic used by the workflow constraints.

Although we have not exhaustively tested with many different
inconsistency types, the results we have obtained so far and the
flexibility of the underlying platform, lead us to believe that
PCV is able to find most types of inconsistencies within and
between security policies and other specifications.

\section{Related work}\label{sec:related}

The security inconsistency problem has been addressed by many
authors. Some solve conflicts within security specifications by
adding implicit rules to incomplete specifications
\cite{Jajodia:1997:UFE,Bertino:1999:CSFW,Li:1999:CSFW}. Others
detect inconsistencies in security properties within workflow
specifications \cite{RBAC97*1,AtlHua96}.

Jajodia {\em et al} \cite{Jajodia:1997:UFE} define a logical
language with ten predicate symbols to express security policies.
Three of them are authorization predicates (dercando, cando, do),
used to define the allowed actions. Although not explicitly, these
predicates define three authorization levels with ``dercando" as
the weakest and ``do" as the strongest. The ``do" predicates are
used to solve conflicts between ``cando" predicates. ``dercando"
predicates are derived from ``cando" predicates, and are
overridden by ``cando" predicates when in conflict.

Another approach to conflict resolution, presented in
\cite{Bertino:1999:CSFW} and \cite{Li:1999:CSFW} uses elements
such as rule authorship authority, rule specificity and rule
recency to prioritize rules. Although simple and natural, this
approach may lead to undesired behavior. It is not uncommon for a
high authority manager to issue a rule which may be overridden by
a low authority manager, or to express a mandatory general rule
which should not be overridden.

Bertino {\em et al} \cite{RBAC97*1} also use constraints to detect
inconsistencies of roles assignment to workflow tasks, and to plan
effective inconsistent-free role assignments. Although the work
is able to model several types of restrictions on role
assignment, namely several forms of separation of duty, it does
not consider any other kind of security or workflow restrictions.

Atluri and Huang \cite{AtlHua96} use a different approach to
detect inconsistencies between security and workflow
specifications. They model security and workflow restrictions as
Petri nets, and state that the safety problem in the
authorization model is equivalent to the reachability problem in
that type of nets. They assume a model where authorization
restrictions are the subset of workflow restrictions that specify
users and roles authorizations.

However, to our knowledge, the general problem of inconsistency
detection on complex security policies, comprising several types
of inconsistency, including inconsistencies with other
specifications, has never been addressed by any author.

\section{Conclusions}\label{sec:conclusion}

We have defined a tool (PCV) to detect inconsistencies within
security policies and between security policies and workflow
specifications. PCV is able to detect several inconsistency types
within security policies defined with the SPL language, which is
able to express complex security polices, and between those
security policies and WPDL workflow specifications.

PCV is easily adapted to each organization's needs by allowing the
definition of other inconsistency types and target specifications.

Currently, our prototype has approximately 300 CHR rules running
over the CHR solver provided with SICstus Prolog
\cite{ercim.sics//T93-01}, and is able to handle all SPL and WPDL
constraints. Some experiences have been performed using
compositions of SPL policies and workflow specifications to
validate the process. Namely, we have tested several workflow
specifications with security policies comprising separation duty,
information flow, and other types of security policies, with
promising results.

This work is part of a security platform which also includes a
security language able to simultaneously express several complex
security policies --the SPL language-- and a compiler which is
able to enforce it efficiently within an event monitor service.

\subsection{Acknowledgments}*

We thank David Matos for reviewing earlier of this article. We
also thank the reviewers of CL2000 workshop on Rule-Based
Constraint Reasoning and Programming for pointing out some
problems on early version of this article.

\bibliographystyle{abbrv}
\bibliography{security,const}

\end{document}